\documentstyle[preprint,aps,psfig,amssymb]{revtex}
\begin{document}

\draft

\title{Effective $Q$-$Q$ interactions in constituent quark models}

\author{L. Ya. Glozman,${}^1$ Z. Papp,${}^2$ W. Plessas,${}^1$
K. Varga,${}^{2,3}$ and R. F. Wagenbrunn${}^1$ }
\address{${}^1$Institute for Theoretical Physics, University of Graz,
A-8010 Graz, Austria\protect\\
${}^2$Institute of Nuclear Research, Hungarian Academy of Sciences,
H-4001 Debrecen, Hungary\protect\\
${}^3$Department of Physics, Niigata University,
Niigata 950-21, Japan}


\maketitle

\begin{abstract}

We study the performance of some recent  potential models suggested as
effective interactions between constituent quarks. In particular, we address
constituent quark models for baryons with hybrid $Q$-$Q$ interactions stemming
from one-gluon plus meson exchanges. Upon recalculating two of such models we
find them to fail in describing the $N$ and $\Delta$ spectra. Our calculations
are based on accurate solutions of the three-quark systems in both a
variational Schr\"odinger and a rigorous Faddeev approach. It is argued that
hybrid \mbox{$Q$-$Q$} interactions encounter difficulties in describing
baryon spectra due to the specific contributions from one-gluon and pion
exchanges together. In contrast, a chiral constituent quark model with a
$Q$-$Q$ interaction solely derived from Goldstone-boson exchange is capable
of providing a unified description of both the $N$ and $\Delta$ spectra
in good agreement with phenomenology.

\end{abstract}
\vspace{1cm}

\newpage

\section {\bf Introduction}

Traditional constituent quark models (CQM) adopted one-gluon exchange (OGE)
\cite{derujula} as the interaction between constituent quarks ($Q$). Over the
years it has become evident that CQM relying solely on OGE $Q$-$Q$ interactions
face some intriguing problems in light-baryon spectroscopy
\cite{glozrisk,gloz}. Most severe are:\\
(i) the wrong level orderings of positive- and negative-parity excitations in
the $N$, $\Delta$, $\Lambda$, and $\Sigma$ spectra;\\
(ii) the missing flavour dependence of the $Q$-$Q$ interaction necessary, e.g.,
for a simultaneous description of the correct level orderings in the $N$ and
$\Lambda$ spectra; and\\
(iii) the strong spin-orbit splittings that are produced by the OGE
interaction but not found in the empirical spectra.\\
All of these effects can be traced back to inadequate symmetry properties
inherent in the OGE interaction.

Beyond CQM with OGE dynamics only, hybrid models advocating in addition
meson-exchange $Q$-$Q$ interactions have been suggested, see, e.g., Ref.
\cite {obukh}. With respect to the $N$ and $\Delta$ spectra, especially
$\pi$ and $\sigma$ exchanges have been introduced to supplement the
interaction between constituent quarks. Sometimes hybrid CQM have also been
employed in constructing baryon-baryon interaction models, see, e.g., Refs.
\cite{obukh,tuebingen}.

Recently two groups, viz. Valcarce, Gonz\'alez, Fern\'andez, and Vento (VGFV)
\cite{valcarce} and Dziembowski, Fabre, and Miller (DFM) \cite{dziembowski}
came up with new versions of hybrid constituent quark models. They presented 
$N$ and $\Delta$ excitation spectra and claimed to have achieved a reasonable
consistent description thereof. In both works a sizeable contribution from the
OGE interaction is maintained. In either case the hyperspherical-harmonics
method (HHM) \cite{ballot,fabre} with several simplifying assumptions was used
to solve the nonrelativistic three-$Q$ problem. We have studied the models of
VGFV and DFM in a detailed manner. In particular, we have checked the
pertinent results for the $N$ and $\Delta$ spectra in two very reliable
ways, namely, by calculating the three-$Q$ systems through\\
(i) a precise solution of the Faddeev equations (via high-rank separable
expansions) \cite{papp} and\\
(ii) a stochastic variational solution of the Schr\"odinger equation
\cite{varga}.\\
Both of these methods have been extensively tested before not only for
three-$Q$ systems like, e.g., for the CQM models of Refs.
\cite{silvestre,despl,glpapl} but also for nuclear and atomic three-body
problems. Existing benchmark results in the literature have been met
accurately, in particular for spin-isospin/flavour dependent forces,
and even a wider class of two-body potentials \cite{papp,varga};
for instance, the stochastic variational method has recently passed successful
tests in three-$N$ (and also four-$N$) bound-state calculations with
realistic $N$-$N$ potentials containing noncentral components of various
spin-spin and tensor types \cite{varga2}. While in all cases our results
for the light-baryon spectra coincide with each other for both of the
above methods (i) and (ii), they are found in striking
disagreement with the results given in the papers of VGFV \cite{valcarce}
and DFM \cite{dziembowski}. For example, the $N$-$\Delta$ splitting is found
to be about 600 MeV with the DFM interaction and even bigger than 2000 MeV
for the VGFV model!

In the following section we give a detailed analysis of the VGFV and DFM models
and outline their shortcomings. We present the corrected spectra and explain in
which respects the physical implications drawn from erroneous results in the
works of VGFV and DFM cannot be relied on.
In Sec. III we discuss a constituent quark model whose effective $Q$-$Q$
interaction is derived from GBE alone \cite{glozrisk}. In the $N$ and $\Delta$
cases the interaction is thus mediated by $\pi$-, $\eta$-, and
$\eta'$-exchanges \cite{glpapl}. Along a modified parametrization of the
corresponding  pseudoscalar meson-exchange potentials in a semirelativistic
framework it is demonstrated that
no OGE (or at most a very weak one) is needed to provide a unified description
of the $N$ and $\Delta$ spectra in close agreement with phenomenology.

\section{Constituent quark models with hybrid $Q$-$Q$ interactions}

Hybrid CQM advocate sizeable contributions from OGE and
employ in addition meson exchanges for the effective interaction between
constituent quarks.

\subsection{The model by Valcarce, Gonz\'alez, Fern\'andez, and Vento}

In the model of VGFV \cite{valcarce} the following $Q$-$Q$ potential is used:
\begin{equation}
\label{vvalc}
V^{VGFV}=V_{OGE}+V_{OPE}+V_{OSE}+V_{conf}.
\end{equation}
The OGE potential is employed without tensor and
spin-orbit parts in the actual calculations \cite{privgonz}, i.e. in the form
\begin{equation}
\label{voge}
V_{OGE}(\vec{r}_{ij})=\frac{1}{4}\alpha_s
\vec{\lambda}_i^C\cdot\vec{\lambda}_j^C\left[\frac{1}{r_{ij}}-
\frac{1}{4m_Q^2}(1+\frac{2}{3}
\vec{\sigma}_i\cdot\vec{\sigma}_j)\frac{e^{-r_{ij}/r_0}}{r_{ij}r_0^2}\right],
\end{equation}
where $\vec{\sigma}_i$ and $\vec{\lambda}_i^C$ are the quark spin and colour
matrices, $r_{ij}$ is the interquark distance, and $m_Q$ the constituent quark
mass; VGFV use $m_Q=313$ MeV. $\alpha_s$ is the effective quark-gluon coupling
constant and $r_0$ a parameter involved in the smearing of the
$\delta$-function. The one-pion-exchange (OPE) potential $V_{OPE}$ is used
in the form
\begin{eqnarray}
\label{vope}
V_{OPE}(\vec{r}_{ij}) &=& \frac {g_{\pi Q}^2}{4\pi} \frac {1}{12 m_Q^2} 
\vec{\tau_i} \cdot \vec{\tau_j}
\frac{\Lambda^2}{\Lambda^2 - \mu_\pi^2}
\left\{\left( \mu_\pi^2 \frac {e^{- \mu_\pi r_{ij}}}{r_{ij}}
- \Lambda^2 \frac {e^{- \Lambda r_{ij}}}{r_{ij}}\right)
\vec{\sigma_i} \cdot \vec{\sigma_j}\right.\nonumber\\
&+&\left[\left(1+\frac{3}{\mu_\pi r_{ij}}+\frac{3}{\mu_\pi^2r_{ij}^2}\right)
\mu_\pi^2 \frac {e^{- \mu_\pi r_{ij}}}{r_{ij}}
\left.-\left(1+\frac{3}{\Lambda r_{ij}}+\frac{3}{\Lambda^2r_{ij}^2}\right)
\Lambda^2 \frac {e^{- \Lambda r_{ij}}}{r_{ij}}\right]{\hat{S}}_{ij}\right\},
\end{eqnarray}
where $g_{\pi Q}^2/4\pi$ is the effective pion-quark coupling constant,
$\mu_\pi=0.7$ fm${}^{-1}$ the pion mass, and $\Lambda$ a parameter
related to the extended $\pi Q$ vertex. 
$\hat{S}_{ij}$ is the tensor operator 
$\hat{S}_{ij}=3(\vec{\sigma_i} \cdot \hat{r}_{ij})
(\vec{\sigma_j} \cdot \hat{r}_{ij})
-\vec{\sigma_i} \cdot \vec{\sigma_j}$.
The one-sigma-exchange (OSE) potential $V_{OSE}$ takes simply the form
\begin{equation}
\label{vose}
V_{OSE}(\vec{r}_{ij}) = - \frac {g_{\pi Q}^2}{4\pi} 
\frac{\Lambda^2}{\Lambda^2 - \mu_\sigma^2}
\left(  \frac {e^{- \mu_\sigma r_{ij}}}{r_{ij}}
-  \frac {e^{- \Lambda r_{ij}}}{r_{ij}}\right)
\end{equation}
with the $\sigma$-mass $\mu_\sigma=3.42$ fm${}^{-1}$.
Finally the confinement potential is assumed in linear form
\begin{equation}
\label{vconf}
V_{conf}(\vec{r}_{ij}) =V_0 + C r_{ij}
\end{equation}
with the strength $C=0.980$ fm${}^{-2}$ and a constant $V_0$ necessary to
adjust the nucleon ground state level to its phenomenological value.
The whole $Q$-$Q$ potential (\ref{vvalc}) involves 6 fitting parameters
($\alpha_s$,  $g_{\pi Q}$, $r_0$, $\Lambda$, $C$, $V_0$) if one considers the
constituent-quark and meson masses as predetermined; the actual values
of the fitting parameters are summarized in Table \ref{vgfvtab}.

VGFV solved the three-$Q$ system along the Schr\"odinger
equation with the HHM \cite{ballot,fabre}. However, only a very restricted
basis set was used for the expansion of the three-$Q$ wave functions (two terms
$K=0,2$ for positive-parity states and one term $K=1$ for negative-parity
states). VGFV published the results shown in Fig. 1a.

We have recalculated the $N$ and $\Delta$ spectra with the VGFV potential in
Eq. (\ref{vvalc}) along both of our methods, i.e., (i) the precise solution
of the Faddeev equations (via high-rank separable expansions) and
(ii) the stochastic variational solution of the Schr\"odinger equation
(as explained in the previous section; for details see also
Refs. \cite{papp,varga}). The results from both calculations agree perfectly
with each other and they are shown in Fig. 1b.

>From the comparison of the spectra in Figs. 1a and b, which have both been
normalized to reproduce the nucleon ground state at 939 MeV (in case of our
calculations a numerical value of $V_0=+488$ MeV is
to be employed for this purpose in the confinement potential
(\ref{vconf})), several differences are evident. Most striking is the unrealistic
$N$-$\Delta$ splitting, which turns out to be larger than 2000 MeV; notice that
the $\Delta$ ground state has a mass of $\approx 3200$ MeV! Also some of
the $N^\ast$ levels, especially in the negative-parity
$\frac{1}{2}^-,\frac{3}{2}^-,\frac{5}{2}^-$ bands, show excitations above
the $N$ ground state by far too large. Furthermore, additional levels (not
given in Ref. \cite{valcarce}) appear. For instance, in the $\frac{1}{2}^-$
band we reproduce the lowest state, which is predominantly $S=\frac{1}{2}$,
$L=1$, at a mass of $\approx 1550$ MeV. The next excitations, whose spin contents
are still $S=\frac{1}{2}$, $L=1$ (with probabilities $\approx 99\%$), are found
between 2000 and 2500 MeV. Only much further above, at $\approx 3500$ MeV,
the $S=\frac{3}{2}$, $L=1$ excitation (marked
by a dagger in Fig. 1b) appears. This is the state to be compared with the
second $\frac{1}{2}^-$ excitation (also marked by a dagger) in the spectrum
given by
VGFV in Fig. 1a. Exactly the same structure is recovered in the $\frac{3}{2}^-$
and $\frac{5}{2}^-$ bands, with even an additional $S=\frac{1}{2}$, $L=3$ state
appearing in the latter. 
In Ref. \cite{valcarce} also the prediction for the Roper resonance
$N(1440)$ is given with a wrong energy. In fact, it lies above 1600 MeV
and thus also above the first negative-parity excitation (see Fig. 1b).
Consequently the problem with the wrong level orderings of positive- and
negative-parity states persists, contrary to the claim made by VGFV.

We conclude that the HHM as used by VGFV
is inadequate to reproduce the correct spectra in the case of the $Q$-$Q$
potential (\ref{vvalc})--(\ref{vconf}) with the parameters chosen in Ref.
\cite{valcarce}. The most important reason is that for certain spin-isospin
channels, specifically for $S_{ij}=T_{ij}=0$, the potentials (\ref{voge})
and (\ref{vope}) become very deep at rather short distances $r_{ij}$. This is
especially due to the OGE potential whose $\delta$-function term
in the central and spin-spin parts involves an
extremely small parameter $r_0$ (see Table \ref{vgfvtab}). If also the OGE
tensor part as given in formula (2) of Ref. \cite{valcarce} was employed,
the three-$Q$ spectrum would not be bounded from below. Even in the case
one uses just the OGE potential as in Eq. (\ref{voge}) above (with such a
small $r_0$ as employed by VGFV) 
great care must be exerted in the expansion of test functions. While
the actual calculation is not so dangerous in the $\Delta$ case (here, 
$S_{ij}=T_{ij}=1$, thus the OGE is repulsive and the OPE is attractive at short
ranges), the $N$ case is rather tricky to compute (here, $S_{ij}=T_{ij}=0$, and
both the OGE and OPE are extremely attractive). In this situation the HHM
result with only a few (two) basis functions is simply not converged. If a
sufficiently rich basis is employed, the $N$ ground state (as well as some of
its excitations) falls down by more than 2000 MeV and additional levels of
spin content $S=\frac{1}{2}$ are found below the $S=\frac{3}{2}$ states in the
$J^P=\frac{1}{2}^-$, $J^P=\frac{3}{2}^-$, and $J^P=\frac{5}{2}^-$ bands.
The convergence of the energies to the values given in Fig. 1b can be followed
in a transparent manner especially in the case of our variational solution of
the Schr\"odinger equation, when gradually increasing the test-function space.
>From the fully converged results we must then conclude that the VGFV model withthe parameter set in Table \ref{vgfvtab} fails completely in describing the
$N$ and $\Delta$ spectra.

\subsection{The model by Dziembowski, Fabre, and Miller}

In the DFM model the effective $Q$-$Q$ interaction
(called ``complete Hamiltonian'' in Ref. \cite{dziembowski}) consists of 
\begin{equation}
\label{vdziemb}
V^{DFM} = V_{OGE} + V_{\chi}+V_{conf}.
\end{equation}
The OGE potential \cite{derujula} was taken in the usual form but again without
spin-orbit and tensor parts --- just as in the case of the VGFV model in Eq.
(\ref{voge}). Also a linear confinement as in Eq. (\ref{vconf}) was
used. The potential $V_\chi$ relating to broken chiral symmetry was
assumed after the suggestion of GBE in Ref. \cite {glozrisk}. In the actual
calculation, DFM took the central part of the pseudoscalar-octet meson-exchange
potential (i.e., for $\pi$ and $\eta$) whose radial dependence was parametrized
as
\begin{equation}
\label{vchi}
V_\chi(\vec{r}_{ij}) = \frac {g_{\pi Q}^2}{4\pi} \frac {1}{12 m_Q^2} 
\vec{\sigma_i}\cdot\vec{\sigma_j}
\vec{\lambda_i}^F \cdot \vec{\lambda_j}^F
\left( \mu_\gamma^2 \frac {e^{- \mu_\gamma r_{ij}}}{r_{ij}}
- \Lambda^2 \frac {e^{- \Lambda r_{ij}}}{r_{ij}}\right),\qquad
\gamma=\pi,\eta.
\end{equation}
Here, $\vec{\lambda_i}^F$ are the Gell-Mann flavour matrices, $m_Q=336$ MeV the
constituent quark mass, $\mu_\gamma$ the meson
masses, and $\Lambda=1/r_0$ a parameter, again from the smearing of the
$\delta$-function. Contrary to VGFV, in the DFM model no $\sigma$-exchange was
employed. Indeed, the effect of this spin- and flavour-independent scalar
interaction can always be incorporated effectively in the confinement    
potential (\ref{vconf}). The fitting parameters of the DFM $Q$-$Q$
interaction are given in Table \ref{dfmtab}. The DFM group solved the three-$Q$
system also along the Schr\"odinger equation with the HHM \cite{ballot,fabre}.
However, they used an even more restricted basis than VGFV, employing only a
single state for both positive- as well as negative-parity states 
and leaving out mixed-symmetry spin-flavour configurations. In case of
their ``complete Hamiltonian'' (i.e. the hybrid model with OGE and one-meson
exchange and the parameter set of Table \ref{dfmtab}) they obtained the results
shown in Fig. 2a.

We have recalculated the DFM model again by both the stochastic variational and
Faddeev methods (as described in the previous section) and found the $N$ and
$\Delta$ spectra as shown in Fig. 2b. For adjusting the $N$ ground state to its
empirical value of 939 MeV we have to take a constant $V_0=-86$ MeV in the
confinement potential (\ref{vconf}). Considerable differences are evident when
comparing Figs. 2a and b. Again, most striking is the huge $N$-$\Delta$ splitting,
which turns out to be more than two times larger than in reality. Also the firstexcited levels in the $J^P=\frac{1}{2}^+,\frac{1}{2}^-,\frac{3}{2}^-$ nucleon
bands differ by 40-50 MeV; the discrepancies are of a similar magnitude in
the $J^P=\frac{3}{2}^+,\frac{5}{2}^+$ states. The higher excitations in both
the positive- and negative-parity bands show unrealistic splittings from the
$N$ ground state ($\gtrsim 900$ MeV). Similarly to the VGFV model additional
states appear in the $J^P=\frac{1}{2}^-$ and $\frac{3}{2}^-$ bands: the
$S=\frac{3}{2}$, $L=1$ states are found only above a second $S=\frac{1}{2}$,
$L=1$ excitation. From our recalculated spectra it is also evident that in the
DFM model the wrong level orderings of positive- and negative-parity excitationspersist. For example, the Roper resonance $N(1440)$ lies above the first
negative-parity excitations $N(1535)$-$N(1520)$; the structure is
analogous among the $\Delta$ levels.

Though the DFM potential (\ref{vdziemb}), being less deep than the VGFV
potential at short distances (due to the larger value of $r_0$), is by far
easier to compute than the $Q$-$Q$ interaction of VGFV,
the HHM still was not applied on a large enough basis in Ref. \cite{dziembowski}.
As a result the $N$ level (and some of its excitations) were not converged.
If in our stochastic variational calculation we restrict the test-function
space in a similar manner as DFM, we can roughly reproduce their results
given in Ref. \cite{dziembowski}. However, as it becomes evident from the
fully converged results in Fig. 2b, the DFM model, with the
parameter set in Table \ref{dfmtab}, grossly fails in describing the $N$
and $\Delta$ spectra. 
 
\section{The GBE constituent quark model}

In this Section we discuss a constituent quark model whose $Q$-$Q$ interaction
relies solely on GBE. We like to demonstrate that in this case a unified
description of the $N$ and $\Delta$ spectra (and, indeed, also of all strange
spectra) can be achieved without the need of advocating a sizeable contribution
from OGE. We construct the CQM in a semirelativistic framework \cite{ckp}, i.e.
taking the kinetic-energy operator in relativistic form
\begin{equation}
\label{relkin}
H_0=\sum_{i=1}^3\sqrt{\vec{p}_i{}^2+m_i^2},
\end{equation}
with $m_i$ the masses and $\vec{p}_i$ the 3-momenta of the constituent quarks.
It is obvious that this intermediate step towards a fully covariant treatment
of the three-$Q$ system (which is still beyond scope at present) already
constitutes an essential improvement over nonrelativistic approaches. Thereby
one can at least avoid several disturbing shortcomings of nonrelativistic CQM.
For example, if one calculates the expectation value of the kinetic energy
in nonrelativistic CQM, such as the hybrid models of the previous
section or the model of Ref. \cite{glpapl}, one finds that the ratio of the
mean velocity $v$ of a constituent quark to the velocity $c$ of light is
$v/c>1$. This clearly indicates that three-$Q$ systems must not be treated in
a nonrelativistic manner. By the use of the relativistic kinetic-energy
operator in Eq. (\ref{relkin}) one avoids such problems a priori and $v/c<1$
is always ensured.

In the GBE constituent quark model, the dynamical part of the
Hamiltonian consists of the $Q$-$Q$ interaction
\begin{equation}
\label{vges}
V=V_\chi+V_{conf},
\end{equation}
where $V_{conf}$ is taken in the usual linear form of Eq. (\ref{vconf}) and
$V_\chi$ is derived from GBE. The latter leads to a spin- and flavour-dependent interaction between constituent quarks $i$ and $j$, whose spin-spin
component is manifested by the sum of the pseudoscalar meson-exchange potentials\cite{glozrisk,glpapl}
\begin{equation}
\label{voct}
V_\chi^{octet}(\vec r_{ij})  =
\left[\sum_{a=1}^3 V_{\pi}(\vec r_{ij}) \lambda_i^a \lambda_j^a
+\sum_{a=4}^7 V_K(\vec r_{ij}) \lambda_i^a \lambda_j^a
+V_{\eta}(\vec r_{ij}) \lambda_i^8 \lambda_j^8\right]
\vec\sigma_i\cdot\vec\sigma_j
\end{equation}
and
\begin{equation}
\label{vsing}
V_\chi^{singlet}(\vec r_{ij}) = \frac {2}{3}\vec\sigma_i\cdot\vec\sigma_j 
V_{\eta'}(\vec r_{ij}),
\end{equation}
with $\vec{\sigma_i}$ and $\vec{\lambda_i}$ the quark spin and flavour matrices,respectively.
In the simplest derivation, when pseudoscalar or pseudovector couplings are
employed in the meson-quark vertices and the boson fields satisfy the linear
Klein-Gordon equation, one obtains, in static approximation,
the well-known meson-exchange potentials
\begin{mathletters}
\label{vgamma}
\begin{equation}
V_\gamma (\vec r_{ij})= \frac{g_8^2}{4\pi}
\frac{1}{12m_im_j}
\left\{\mu_\gamma^2\frac{e^{-\mu_\gamma r_{ij}}}{ r_{ij}}-
4\pi\delta (\vec r_{ij})\right\},
\qquad\gamma=\pi,K,\eta
\end{equation}
\begin{equation}
V_{\eta'} (\vec r_{ij})= \frac{g_0^2}{4\pi}
\frac{1}{12m_im_j}
\left\{\mu_{\eta'}^2\frac{e^{-\mu_{\eta'} r_{ij}}}{ r_{ij}}-
4\pi\delta (\vec r_{ij})\right\},
\end{equation}
\end{mathletters}
with $\mu_\gamma$ being the meson masses ($\gamma=\pi,K,\eta,\eta'$). While
the octet of pseudoscalar mesons may be considered as the manifestation of the
Goldstone bosons, the singlet $\eta'$ is a priori no Goldstone boson due to the
axial anomaly. However, in the large-$N_C$ limit, the axial anomaly would
vanish \cite{witten} and the $\eta'$ become the ninth Goldstone boson of the
spontaneously broken $U(3)_R\times U(3)_L$ chiral symmetry \cite{cole}.
Under real conditions with $N_C=3$, we may thus consider the contribution from
$\eta'$ exchange but with a coupling different from the octet mesons (see the
discussion later on). The potentials (\ref{vgamma}) are strictly applicable
only for pointlike particles. Since one deals with structured particles
(constituent quarks and pseudoscalar mesons) of finite extension, one must smearout the $\delta$-function. The corresponding term turns out to be of crucial
importance for the effective $Q$-$Q$ interaction in baryons: it has a sign
appropriate to reproduce the level splittings and it dominates over the Yukawa
part towards shorter distances.

There are several choices of smearing the $\delta$-function in Eqs.
(\ref{vgamma}). A reasonable constraint is that the volume integral of the
potential should vanish --- as it is required for pseudoscalar exchange
interactions with finite-mass bosons (mesons); for pure Goldstone bosons of
vanishing mass this constraint would not apply, however.
In Ref. \cite{glpapl} we have employed the Gaussian-type smearing of the
$\delta$-function
\begin{equation}
4\pi\delta(\vec{r}_{ij})\to\frac{4}{\sqrt{\pi}}\alpha^3
e^{-\alpha^2(r_{ij}-r_0)^2},
\end{equation}
which involves two parameters corresponding to the position ($r_0$) and the
breadth ($\alpha$) of a bell-shaped curve. In this case the volume integral of
the chiral potential $V_\chi$ does not vanish.

We have found \cite{daphce} that a Yukawa-type smearing like 
\begin{equation} 
\label{yuk}
4\pi\delta(\vec{r}_{ij})\to \Lambda_\gamma^2\frac{e^{-\Lambda_\gamma r_{ij}}}
{r_{ij}},\qquad \gamma=\pi,K,\eta,\eta'
\end{equation}
involving the ``cut-off'' parameters $\Lambda_\gamma$, works equally well for
the effective $Q$-$Q$ interaction in baryon spectra. With the $\delta$-function
representation (\ref{yuk}) the volume integral of $V_\chi$
vanishes, thus meeting the requirement mentioned above.

The $\delta$-function representation (\ref{yuk}) relates to the finite
extension of the meson-quark vertices. If we employ the
phenomenological values for the different meson masses $\mu_\gamma$, we should
also allow for a different cut-off parameter $\Lambda_\gamma$ corresponding to
each meson exchange: with a larger meson mass also $\Lambda$ should increase.
Otherwise, the meson-exchange interactions (\ref{vgamma}) could receive
unwarranted contributions from certain meson exchanges (e.g., it could become
effectively attractive instead of
repulsive or vice versa at short distances). In our attempt to keep the number
of free parameters as small as possible, we have avoided fitting each
individual cut-off parameter $\Lambda_\gamma$, however. Instead, we have
succeeded in describing their dependence on the meson mass via the linear
scaling prescription
\begin{equation}
\Lambda_\gamma=\Lambda_0+\kappa\mu_\gamma,
\end{equation}
which involves only the two parameters $\Lambda_0$ and $\kappa$. Their numericalvalues have been determined from a fit to the baryon spectra.

In Eq. (\ref{vgamma}a) we have foreseen a single octet meson-quark coupling
constant $g_8^2/4\pi$. Indeed, in the chiral limit there is only one coupling
constant $g_8^2/4\pi$ for all Goldstone bosons. Due to explicit chiral symmetry
breaking the coupling constants for $\pi$, $K$, and $\eta$ may become
different. Nevertheless, due to the present lack of firm insight and in order
to prevent a proliferation of free parameters, we rely at this stage on a
universal coupling constant $g_8^2/4\pi$, which we assume to be equal to the
pion-quark coupling constant $g_{\pi Q}^2/4\pi$. The latter may be deduced from
the pion-nucleon coupling using the Goldberger-Treiman relations for both the
pion-quark and pion-nucleon vertices:
\begin{equation}
\frac{g_8^2}{4\pi}=\frac{g_{\pi Q}^2}{4\pi}=
\left(\frac{g_Q^A}{g_N^A}\right)^2\left(\frac{m_Q}{m_N}\right)^2
\frac{g_{\pi N}^2}{4\pi}.
\end{equation}
Here, $g_Q^A$ and $g_N^A$ are the quark and nucleon axial coupling constants,
$m_Q$ and $m_N$ the light-quark resp. nucleon masses, and
$g_{\pi N}^2/4\pi=14.2$, the phenomenological $\pi N$ coupling constant
\cite{ericson}. The nucleon and quark axial coupling constants can be related
by the ratio
\begin{equation}
\frac{g_Q^A}{g_N^A}=\frac{3}{5}=0.6.
\end{equation}
If one takes a constituent-quark mass of $m_Q=340$ MeV (as it is
suggested from nucleon magnetic-moment studies) one can deduce the $\pi
Q$ coupling constant of the size $g_{\pi Q}^2/4\pi=0.67$.
Thus we fix the octet meson-quark coupling to this value, as it was done
already in Ref. \cite{glpapl}.

We remark, however, that the $\pi Q$ coupling constant could also assume a
value slightly different from $0.67$. For instance, if one takes for the $Q$
axial coupling constant $g_Q^A=1$ (as it
should be in the large-$N_C$ limit \cite{weinberg1}) and employs for the $N$
axial coupling constant its phenomenological value \cite{ericson} $g_N^A=1.25$,
then one obtains the ratio 
\begin{equation}
\frac{g_Q^A}{g_N^A}=0.8.
\end{equation}
Thus one would end up with a pion-quark coupling constant $g_{\pi Q}^2/4\pi=1.19$.
If one took into account that $1/N_C$ corrections would reduce the
$Q$ axial coupling constant to $g_Q^A=0.87$ \cite{weinberg2},
then
\begin{equation}
\frac{g_Q^A}{g_N^A}=0.7,
\end{equation}
and consequently $g_{\pi Q}/4\pi=0.91$. Therefore we may ultimately expect the
$\pi Q$ coupling constant within the interval
$0.67\lesssim g_{\pi Q}^2/4\pi\lesssim 1.19$ and this is even dependent on the
value of the constituent-quark mass adopted. Still, at this stage, we need not
vary the size of the octet meson-quark coupling constant and simply fix it to
the ``canonical'' value of $\approx 0.67$.

Due to the special character of the singlet $\eta'$ meson (see the discussion
above), its coupling was allowed to deviate from the universal octet coupling.
In the actual fit, the ratio $(g_0/g_8)^2$ turns out larger than 1. However,
this is in line with results deduced in Refs.
\cite{chengli1,chengli2} from the Gottfried sum-rule violation, the
$\bar{u}/\bar{d}$ ratio, as well as the spin content of the
nucleon.

In the semirelativistic GBE model we have a confinement strength (of the linear
confining potential, Eq. (\ref{vconf})) of the order of $C\approx 2.3$
fm${}^{-2}$. We note that this value appears to be quite realistic, as it is
consistent both with Regge slopes and also the string tension extracted in
lattice QCD \cite{creutz}. The size of the confinement strength required in
our semirelativistic GBE model represents another comfortable improvement over
nonrelativistic CQM where a much smaller value of $C$ must be chosen (see
the examples of the previous section or Refs. \cite{glpapl,daphce}).

For the whole $Q$-$Q$ potential in Eq. (\ref{vges}) our model now involves
a total of five
free parameters whose numerical values are given in Table \ref{yuktab}. We remark,
however, that this is only one fit out of a possible set of others that lead to
a similar quality in the description of the baryon spectra. For instance, we
have found that one
could increase the octet coupling constant $g_8^2/4\pi$ and compensate the change
by readjusting the other parameters, particularly the cut-offs $\Lambda_\gamma$.Similarly, the ratio $(g_0/g_8)^2$ could be lowered, and even the
constituent-quark masses may be chosen differently. Retuning the other parameters,
including the confinement strength $C$, one can within certain limits always
obtain a fit of comparable quality.

The GBE model with the semirelativistic three-$Q$ Hamiltonian relying on
Eqs. (\ref{relkin}) and (\ref{vges}) was solved along the stochastical
variational method \cite{varga} in momentum space. The convergence was
carefully tested, especially with respect to the $N$ ground state and the
``dangerous'' levels in the excitation spectra.

In Fig. 3 we present the ground-state and excitation levels of the $N$ and
$\Delta$ spectra as produced by the semirelativistic GBE constituent quark
model with the $Q$-$Q$ potential from Eqs. (\ref{vges})--(\ref{vgamma}), the
Yukawa-type representation  (\ref{yuk}) of the $\delta$-function and the
parameter set given in Table \ref{yuktab}. The light-$Q$ masses were taken as
$m_u=m_d=m_Q=340$ MeV, and for the meson masses $\mu_\gamma$ the
phenomenological values were employed.

Tensor force effects are not yet included in these results, as is evident
from the absence of any fine-structure splittings in the theoretical
$LS$-multiplets. It is clear, however, that tensor forces can at most play a
subordinate role in the $N$ and $\Delta$ spectra. First of all, this follows
from the smallness of the level splittings of corresponding $LS$-multiplets in
the experimental spectra. Second, from a first numerical estimation of the
influence of the pseudoscalar meson-exchange tensor forces prevailing in the
GBE model we have found only small effects on the baryon states.

>From the results of Fig. 3 it becomes evident that within the GBE
constituent quark model a unified description of both
the $N$ and $\Delta$ spectra is achieved in good agreement with phenomenology.
Practically all levels are found in their ``experimental boxes'' and
specifically the orderings of positive- and negative-parity excitations are
correct. 

\section{Discussion}

As already outlined before \cite{glozrisk,glpapl}, the successes of the GBE
constituent quark model are due to the specific spin-flavour symmetry inherent
in the chiral
potential $V_\chi$ in Eqs. (\ref{voct}) and (\ref{vsing}). It produces a
level structure well adjusted to the experimental data. This is not
only true in the $N$ and $\Delta$ cases but holds also for strange baryons
\cite{daphce,relativistic}.
With the possible exception of the empirically large
$\Lambda(1405)$-$\Lambda(1520)$ splitting, which remains
unexplained at this stage, all octet/decuplet states together with their
excitations can be described in a unified manner.

>From the results of the GBE constituent quark model exemplified in the previoussection we also learn that no (additional) OGE is needed for the reproduction
of the baryon spectra, contrary to the claims made for the hybrid models, e.g.,
in Refs. \cite{valcarce} and \cite{dziembowski}. Still, one could accommodate a
certain contribution from OGE but only with a rather small effect from the
colour-magnetic interaction. As soon as the colour-magnetic forces become
strong, like it is the case in the hybrid models discussed in Sec. II, 
one faces the traditional difficulties in the light-baryon spectra. If in
a meson-exchange plus OGE interaction with a sizeable contribution from the 
colour-magnetic interaction  the $N$-$\Delta$ splitting (to which both types of
exchanges contribute in the same manner) is described correctly, the correct
ordering of positive- and negative-parity levels is not achieved, because the
meson-exchange contribution turns out simply too weak. Only a bigger influence
from meson (pion) exchange would allow the first positive-parity excitations in
the $N$ spectrum to fall below the first negative-parity
excitations. If, on the other hand, the meson exchange is made strong enough to
achieve this level inversion (as demanded by phenomenology), the colour-magneticcontribution must remain very small, practically negligible, in order not to
spoil a correct $N$-$\Delta$ splitting. Thus it is quite naturally suggested
by the experimental level structure that OGE can at most play a subordinate
role in CQM for light baryons.

Finally we note that a nonrelativistic treatment of three-$Q$ systems cannot
really work. Evidently, light (and strange) three-$Q$ systems already bear
large relativistic effects of kinematical origin. They call at least for the
kinetic energy part of the full Hamiltonian to be employed in relativistic
form, as was done for the GBE model in Sec. III. Otherwise the lack of taking
into account (kinematical) relativistic effects explicitly will be compensated
within the $Q$-$Q$ potential parameters in the fit to the baryon spectra.
Consequently, in this procedure the model parameters will assume rather
unrealistic values, e.g., the confinement strength would turn out to be only
half of the one resulting for the string tension from lattice QCD and the one
needed to fit Regge slopes. As a result, any nonrelativistic CQM can at most be
considered as a parametrization of the baryon energy levels, rather than as a
dynamical model for light three-$Q$ systems. Certainly it will not prove
acceptable for future applications such as the description of electromagnetic
form factors, hadronic decays, and other dynamical observables that are
determined by the behaviour of the baryon wave functions and are generally
much influenced by $Q$-$Q$ potential parameters.

\section*{Acknowledgment}
The authors acknowledge clarifying correspondence with P. Gonz\'alez,
A. Valcarce, and V. Vento as well as discussions with G.A. Miller.
They have also profited from valuable discussions with D.O. Riska as well
as B.L.G. Bakker, V.B. Mandelzweig, and  M. Rosina. The work was partially
supported by the Austrian-Hungarian Scientific-Technical Cooperation,
within project A23 resp. 17/95, and by the Paul-Urban Foundation.

\begin{figure}
\vspace{1cm}
\psfig{file=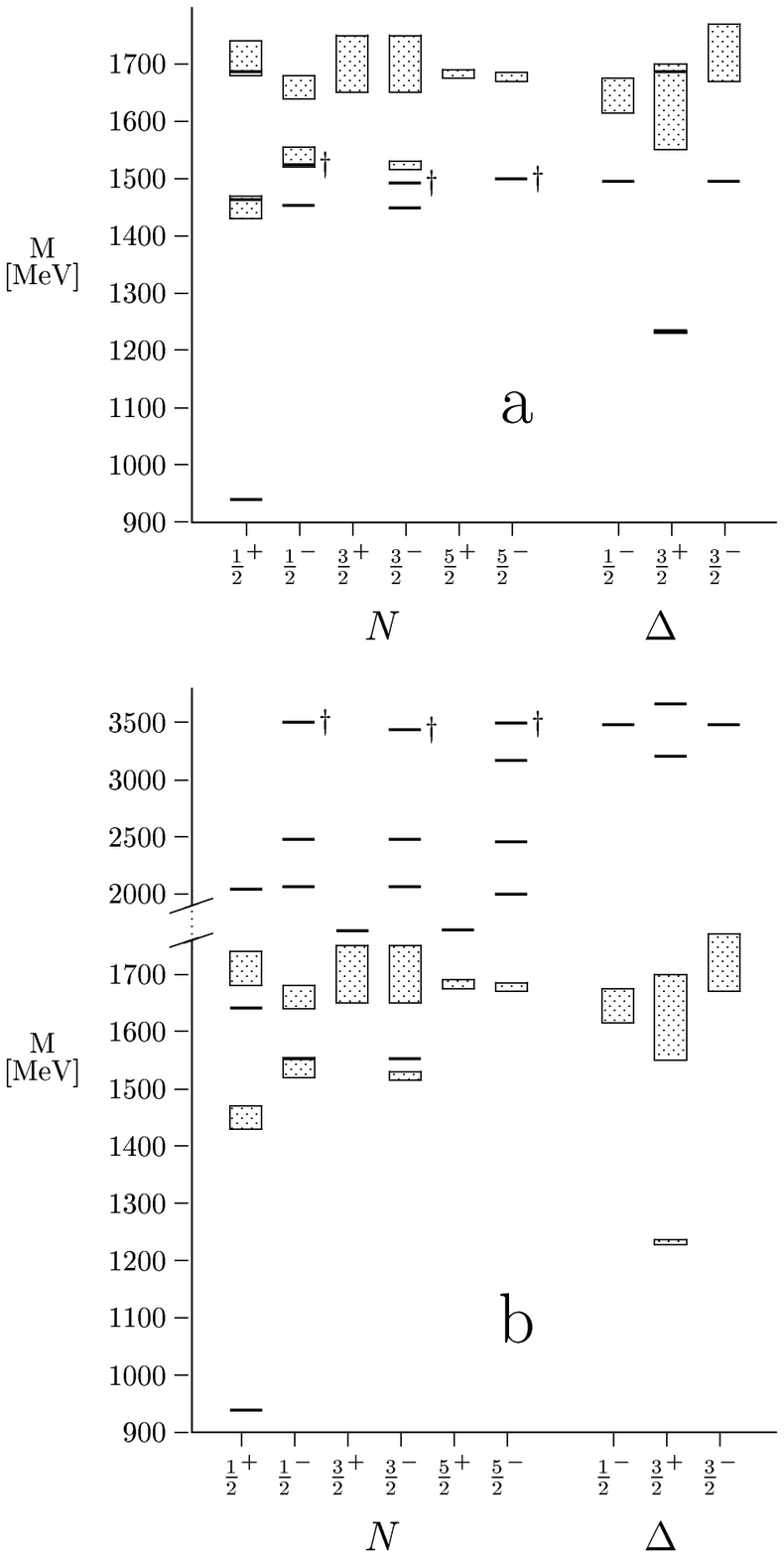,width=10.5cm}
\caption{$N$ and $\Delta$ spectra in case of the model VGFV: a) as given in
Ref. \protect\cite{valcarce}, b) recalculated. In both cases the nucleon mass
is ``normalized'' to its phenomenological value of 939 MeV. In the recalculated
spectra the $N$-$\Delta$ mass difference is 2270 MeV.
The negative-parity states
marked by a dagger correspond to the spin content of predominantly
$S=\frac{3}{2}$.}
\end{figure}

\begin{figure}
\psfig{file=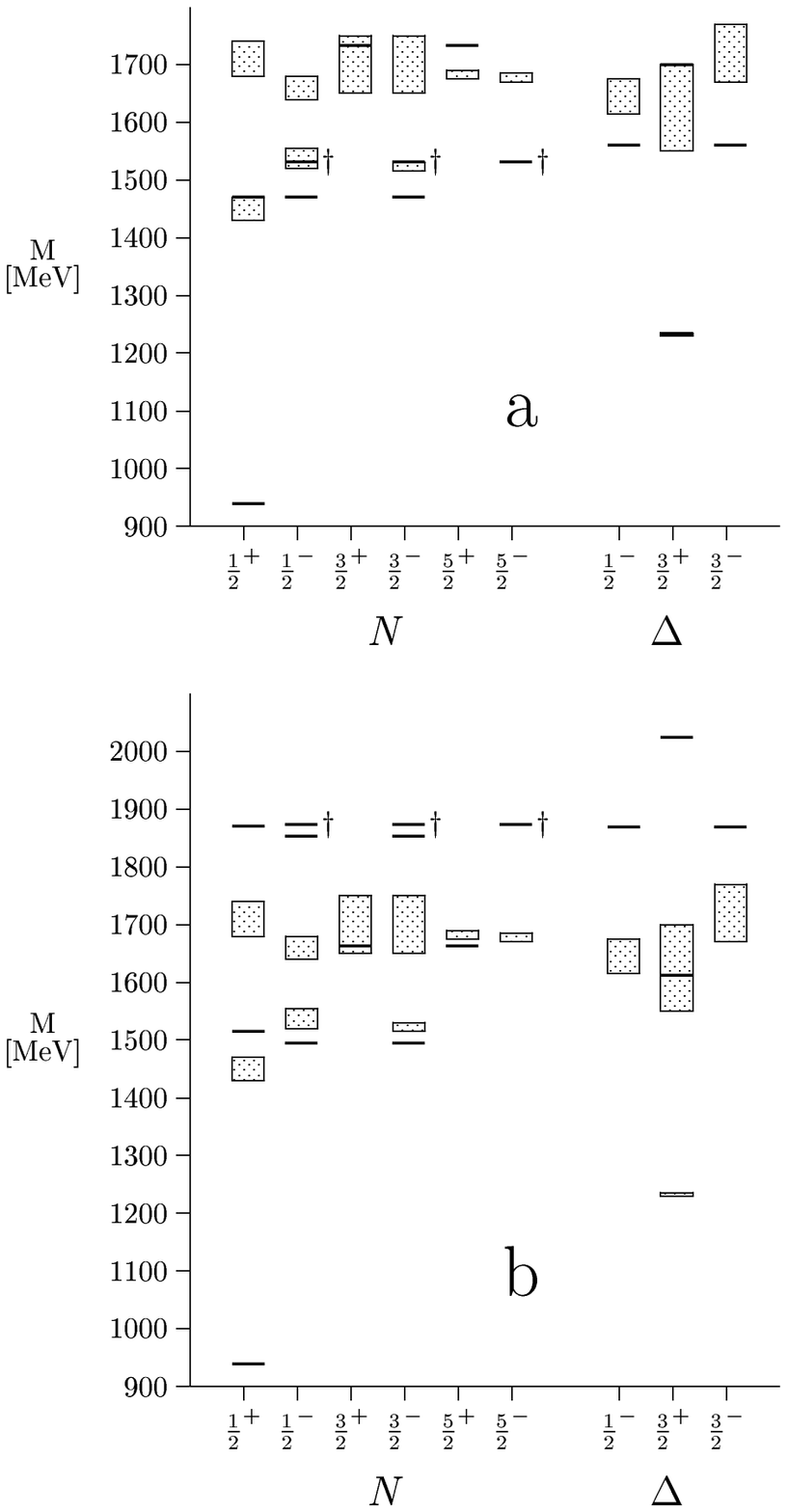,width=10.5cm}
\caption{$N$ and $\Delta$ spectra in case of the model DFM: a) as given in
Ref. \protect\cite{dziembowski}, b) recalculated. In both cases the nucleon
mass is ``normalized'' to its phenomenological value of 939 MeV. In the
recalculated spectra the $N$-$\Delta$ mass difference is 670 MeV. The
negative-parity states marked by a dagger correspond to the spin content 
$S=\frac{3}{2}$.}
\end{figure}

\begin{figure}
\psfig{file=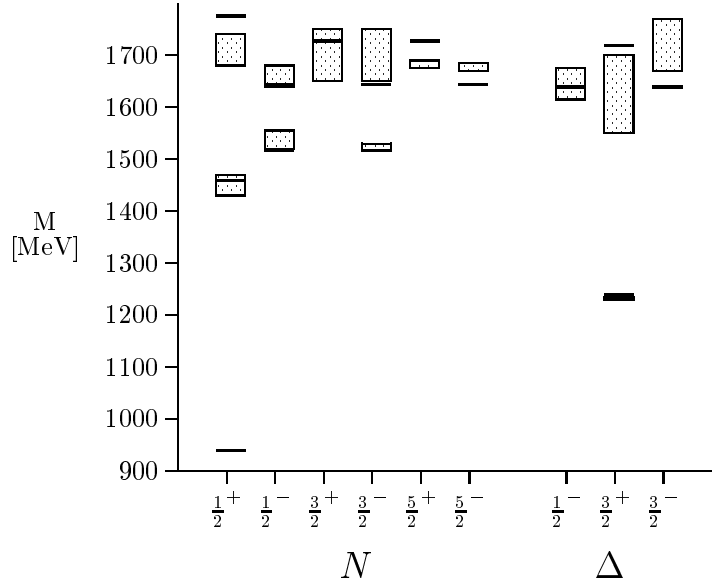,width=10.5cm}
\vspace{1cm}
\caption{Energy levels of the 14 lowest light-baryon states with total angular
momentum and parity $J^P$ in case of the semirelativistic GBE constituent quark
model with parameters as given in Table \protect\ref{yuktab}. The
shadowed boxes represent the experimental values with their uncertainties; the
$N$ and $\Delta$ ground-state levels coincide with the experimental values.}
\end{figure}

\begin{table}
\caption{Numerical values of fitting parameters of the VGFV $Q$-$Q$
interaction. Three of the parameters, namely, $\alpha_s$, $g_{\pi Q}^2/4\pi$,
and $\Lambda$ were taken from the hybrid $N$-$N$ interaction model in Ref.
\protect\cite{valcarce2}.}
\label{vgfvtab}
\begin{tabular}{cccccc}
$\alpha_s$&$g_{\pi Q}^2/4\pi$& $C$ $\rm[fm^{-2}]$&$V_0$ $\rm[MeV]$&$r_0$
$\rm[fm]$&$\Lambda$ $\rm[fm^{-1}]$\\ \hline
0.485&0.545&0.980\tablenote{There is a misprint in the value of the confinement
strength in Ref. \cite{valcarce}. The correct value given here was communicated
to us by the Salamanca-Valencia group \cite{privvalc}.}&
$+488$\tablenote{The constant
$V_0$ in the confinement potential, whose value is not quoted in Ref.
\cite{valcarce}, must be taken of this magnitude in order to adjust the nucleon
ground state in our (converged) calculation.}&
0.0367&4.2
\end{tabular}
\end{table}
\vfill
\begin{table}
\caption{Numerical values of fitting parameters of the DFM $Q$-$Q$ interaction.
The cut-off parameter $\Lambda$ in the meson-exchange potentials
(\protect\ref{vchi}) was chosen to be the same as in the OGE part, i.e.
$\Lambda=1/r_0$.}
\label{dfmtab}
\begin{tabular}{ccccc}
$\alpha_s$&$g_{\pi Q}^2/4\pi$& $C$ $\rm[fm^{-2}]$&$V_0$ $\rm[MeV]$&$r_0$
$\rm[fm]$\\ \hline
0.35&1.15&1.014&$-86$\tablenote{The constant $V_0$ in the confinement potential,whose value is not quoted in Ref. \cite{dziembowski}, must be taken of this
magnitude in order to adjust the nucleon ground state in our (converged)
calculation.}&0.238
\end{tabular}
\end{table}
\vfill
\begin{table}
\caption{Numerical values of fitting parameters of the semirelativistic GBE
$Q$-$Q$ interaction. The octet meson-quark coupling constant was fixed to
$g_8^2/4\pi=0.67$.}
\label{yuktab}
\begin{tabular}{ccccc}
$(g_0/g_8)^2$&$C$ $\rm[fm^{-2}]$&$V_0$ $\rm[MeV]$&
$\Lambda_0$ $\rm[fm^{-1}]$&$\kappa$\\ \hline
1.34&2.33&$-416$&2.87&0.81
\end{tabular}
\end{table}

\end{document}